\documentclass[11pt,a4paper]{article}
\usepackage[utf8]{inputenc}
\usepackage{amsmath}
\usepackage{amsfonts}
\usepackage{amssymb}
\usepackage{graphicx}
\usepackage{microtype}
\usepackage{relsize}
\usepackage{xcolor}
\usepackage[sc]{mathpazo}
\usepackage[euler-hat-accent]{eulervm} %euler-digits,
\newcommand{\bA}{\alpha_A}

\newcommand{\dM}{\delta_M}
\newcommand{\dA}{\delta_A}
\newcommand{\dS}{\delta_S}

\newcommand{\kM}{k_M}
\newcommand{\kS}{k_S}

\newcommand{\cMA}{C_{MA}}

\newcommand{\Mr}{M_r}
\newcommand{\Ml}{M_l}
\newcommand{\R}{R}
\newcommand{\Le}{L}
\newcommand{\Br}{B_r}

\newcommand{\Bl}{B_l}

\newcommand{\aMr}{\alpha_{M_r}}
\newcommand{\aMl}{\alpha_{M_l}}
\newcommand{\bR}{\alpha_{R}}%{\beta_{R}}
\newcommand{\bLe}{\alpha_{L}}%{\beta_{Le}}
\newcommand{\dMr}{\delta_{M_r}}
\newcommand{\dMl}{\delta_{M_l}}
\newcommand{\dR}{\delta_{R}}
\newcommand{\dLe}{\delta_{L}}
\newcommand{\kRp}{k_r^+}
\newcommand{\kLp}{k_l^+}
\newcommand{\kRm}{k_r^-}
\newcommand{\kLm}{k_l^-}
\newcommand{\cp}{8 }
\newcommand{\realisations}{2000 }

\renewcommand{\dot}{\partial_t}

\makeatletter
\renewcommand{\fnum@figure}{Fig. S\thefigure}
\renewcommand{\fnum@table}{Table~S\thetable}
\makeatother

\newcommand{\todo}[1]{}

\title{Supplementary Information (Theory):\\ Theoretical Model for ColicinE2 Expression Including the Additional CsrA Sequestering Element ssDNA }
\date{}

\pdfoutput=1
\usepackage{hyperref}
\hypersetup{
  pdfinfo={
    Title={ CsrA and its regulators control the time-point of ColicinE2 release in Escherichia coli},
    Author={Alexandra Götz, Matthias Lechner, Andreas Mader, Benedikt von Bronk, Erwin Frey and Madeleine Opitz},
    Subject={},
    Keywords={CsrA controls the time-point of ColicinE2 release ssDNA, bacteriocin, regulation, rolling circle replication, plasmid copy number}
  }
}

\usepackage{tikz}

\begin{document}

\pagenumbering{gobble} %disables page numbering in the following
\cleardoublepage
\begin{tikzpicture}[remember picture,overlay,shift={(current page.center)}]
\node[inner sep=0pt] at (0,0) {\includegraphics[width=\paperwidth,page=1]{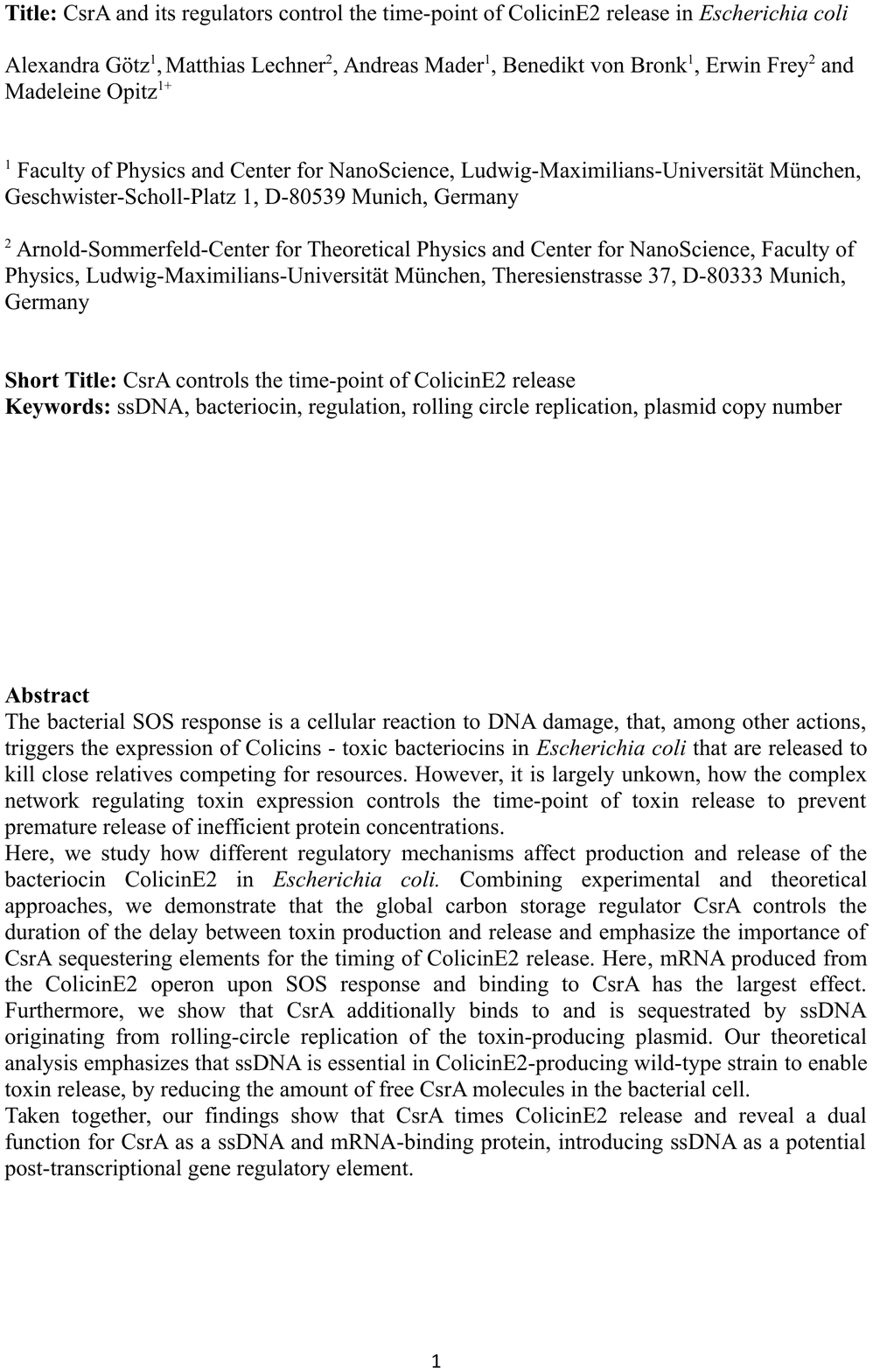}};
\end{tikzpicture}

\cleardoublepage
\begin{tikzpicture}[remember picture,overlay,shift={(current page.center)}]
\node[inner sep=0pt] at (0,0) {\includegraphics[width=\paperwidth,page=2]{arxiv.pdf}};
\end{tikzpicture}

\cleardoublepage
\begin{tikzpicture}[remember picture,overlay,shift={(current page.center)}]
\node[inner sep=0pt] at (0,0) {\includegraphics[width=\paperwidth,page=3]{arxiv.pdf}};
\end{tikzpicture}

\cleardoublepage
\begin{tikzpicture}[remember picture,overlay,shift={(current page.center)}]
\node[inner sep=0pt] at (0,0) {\includegraphics[width=\paperwidth,page=4]{arxiv.pdf}};
\end{tikzpicture}

\cleardoublepage
\begin{tikzpicture}[remember picture,overlay,shift={(current page.center)}]
\node[inner sep=0pt] at (0,0) {\includegraphics[width=\paperwidth,page=5]{arxiv.pdf}};
\end{tikzpicture}

\cleardoublepage
\begin{tikzpicture}[remember picture,overlay,shift={(current page.center)}]
\node[inner sep=0pt] at (0,0) {\includegraphics[width=\paperwidth,page=6]{arxiv.pdf}};
\end{tikzpicture}

\cleardoublepage
\begin{tikzpicture}[remember picture,overlay,shift={(current page.center)}]
\node[inner sep=0pt] at (0,0) {\includegraphics[width=\paperwidth,page=7]{arxiv.pdf}};
\end{tikzpicture}

\cleardoublepage
\begin{tikzpicture}[remember picture,overlay,shift={(current page.center)}]
\node[inner sep=0pt] at (0,0) {\includegraphics[width=\paperwidth,page=8]{arxiv.pdf}};
\end{tikzpicture}

\cleardoublepage
\begin{tikzpicture}[remember picture,overlay,shift={(current page.center)}]
\node[inner sep=0pt] at (0,0) {\includegraphics[width=\paperwidth,page=9]{arxiv.pdf}};
\end{tikzpicture}

\cleardoublepage
\begin{tikzpicture}[remember picture,overlay,shift={(current page.center)}]
\node[inner sep=0pt] at (0,0) {\includegraphics[width=\paperwidth,page=10]{arxiv.pdf}};
\end{tikzpicture}

\cleardoublepage
\begin{tikzpicture}[remember picture,overlay,shift={(current page.center)}]
\node[inner sep=0pt] at (0,0) {\includegraphics[width=\paperwidth,page=11]{arxiv.pdf}};
\end{tikzpicture}

\cleardoublepage
\begin{tikzpicture}[remember picture,overlay,shift={(current page.center)}]
\node[inner sep=0pt] at (0,0) {\includegraphics[width=\paperwidth,page=12]{arxiv.pdf}};
\end{tikzpicture}

\cleardoublepage
\begin{tikzpicture}[remember picture,overlay,shift={(current page.center)}]
\node[inner sep=0pt] at (0,0) {\includegraphics[width=\paperwidth,page=13]{arxiv.pdf}};
\end{tikzpicture}

\cleardoublepage
\begin{tikzpicture}[remember picture,overlay,shift={(current page.center)}]
\node[inner sep=0pt] at (0,0) {\includegraphics[width=\paperwidth,page=14]{arxiv.pdf}};
\end{tikzpicture}

\cleardoublepage
\begin{tikzpicture}[remember picture,overlay,shift={(current page.center)}]
\node[inner sep=0pt] at (0,0) {\includegraphics[width=\paperwidth,page=15]{arxiv.pdf}};
\end{tikzpicture}

\cleardoublepage
\begin{tikzpicture}[remember picture,overlay,shift={(current page.center)}]
\node[inner sep=0pt] at (0,0) {\includegraphics[width=\paperwidth,page=16]{arxiv.pdf}};
\end{tikzpicture}

\cleardoublepage
\begin{tikzpicture}[remember picture,overlay,shift={(current page.center)}]
\node[inner sep=0pt] at (0,0) {\includegraphics[width=\paperwidth,page=17]{arxiv.pdf}};
\end{tikzpicture}

\cleardoublepage
\begin{tikzpicture}[remember picture,overlay,shift={(current page.center)}]
\node[inner sep=0pt] at (0,0) {\includegraphics[width=\paperwidth,page=18]{arxiv.pdf}};
\end{tikzpicture}

\cleardoublepage
\begin{tikzpicture}[remember picture,overlay,shift={(current page.center)}]
\node[inner sep=0pt] at (0,0) {\includegraphics[width=\paperwidth,page=1]{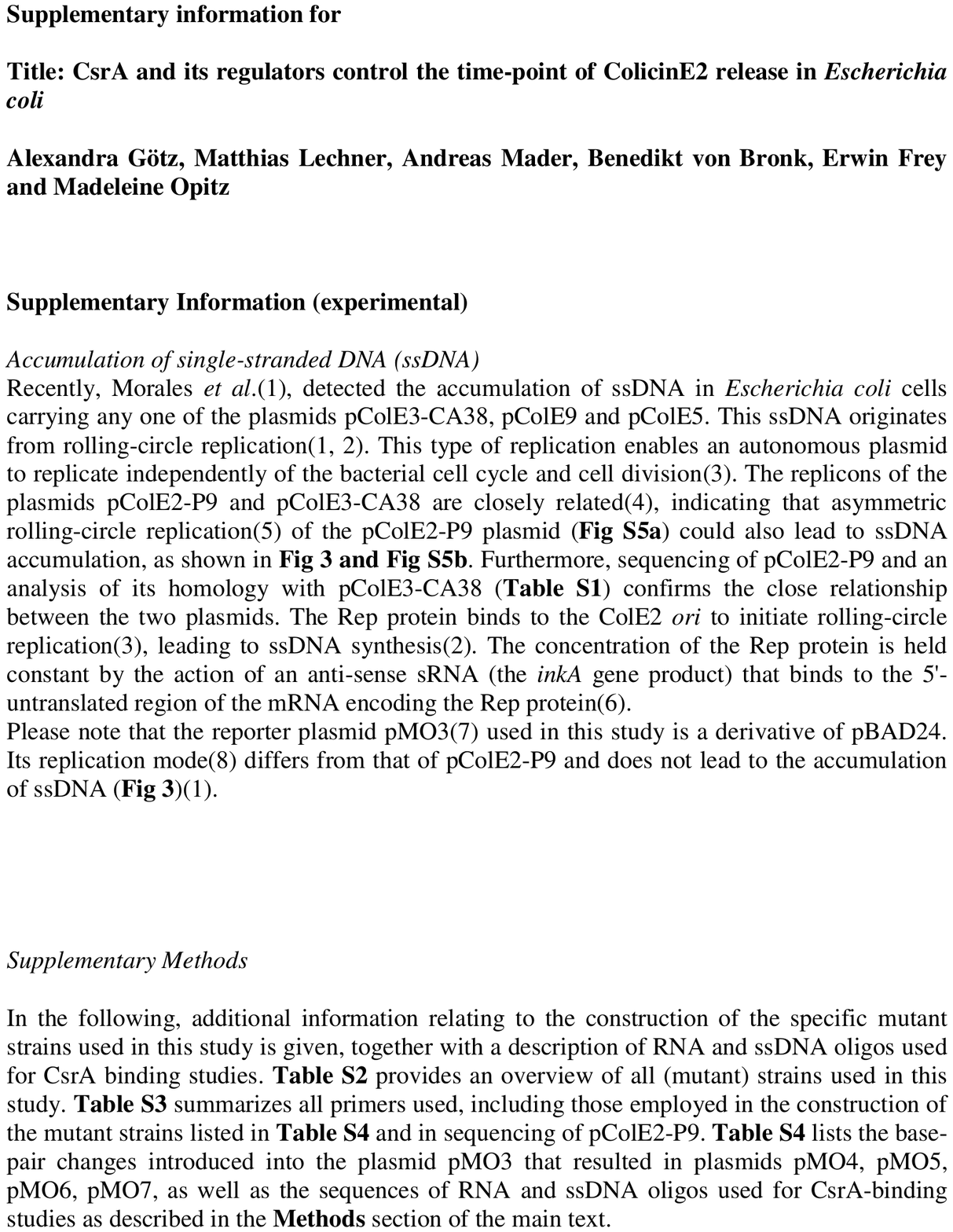}};
\end{tikzpicture}
\cleardoublepage
\begin{tikzpicture}[remember picture,overlay,shift={(current page.center)}]
\node[inner sep=0pt] at (0,0) {\includegraphics[width=\paperwidth,page=2]{arxiv_SI.pdf}};
\end{tikzpicture}
\cleardoublepage
\begin{tikzpicture}[remember picture,overlay,shift={(current page.center)}]
\node[inner sep=0pt] at (0,0) {\includegraphics[width=\paperwidth,page=3]{arxiv_SI.pdf}};
\end{tikzpicture}
\cleardoublepage
\begin{tikzpicture}[remember picture,overlay,shift={(current page.center)}]
\node[inner sep=0pt] at (0,0) {\includegraphics[width=\paperwidth,page=4]{arxiv_SI.pdf}};
\end{tikzpicture}
\cleardoublepage
\begin{tikzpicture}[remember picture,overlay,shift={(current page.center)}]
\node[inner sep=0pt] at (0,0) {\includegraphics[width=\paperwidth,page=5]{arxiv_SI.pdf}};
\end{tikzpicture}
\cleardoublepage
\begin{tikzpicture}[remember picture,overlay,shift={(current page.center)}]
\node[inner sep=0pt] at (0,0) {\includegraphics[width=\paperwidth,page=6]{arxiv_SI.pdf}};
\end{tikzpicture}
\cleardoublepage
\begin{tikzpicture}[remember picture,overlay,shift={(current page.center)}]
\node[inner sep=0pt] at (0,0) {\includegraphics[width=\paperwidth,page=7]{arxiv_SI.pdf}};
\end{tikzpicture}
\cleardoublepage
\begin{tikzpicture}[remember picture,overlay,shift={(current page.center)}]
\node[inner sep=0pt] at (0,0) {\includegraphics[width=\paperwidth,page=8]{arxiv_SI.pdf}};
\end{tikzpicture}
\cleardoublepage
\begin{tikzpicture}[remember picture,overlay,shift={(current page.center)}]
\node[inner sep=0pt] at (0,0) {\includegraphics[width=\paperwidth,page=9]{arxiv_SI.pdf}};
\end{tikzpicture}
\cleardoublepage
\begin{tikzpicture}[remember picture,overlay,shift={(current page.center)}]
\node[inner sep=0pt] at (0,0) {\includegraphics[width=\paperwidth,page=10]{arxiv_SI.pdf}};
\end{tikzpicture}
\cleardoublepage
\begin{tikzpicture}[remember picture,overlay,shift={(current page.center)}]
\node[inner sep=0pt] at (0,0) {\includegraphics[width=\paperwidth,page=11]{arxiv_SI.pdf}};
\end{tikzpicture}
\cleardoublepage
\begin{tikzpicture}[remember picture,overlay,shift={(current page.center)}]
\node[inner sep=0pt] at (0,0) {\includegraphics[width=\paperwidth,page=12]{arxiv_SI.pdf}};
\end{tikzpicture}
\cleardoublepage
\begin{tikzpicture}[remember picture,overlay,shift={(current page.center)}]
\node[inner sep=0pt] at (0,0) {\includegraphics[width=\paperwidth,page=13]{arxiv_SI.pdf}};
\end{tikzpicture}
\cleardoublepage
\begin{tikzpicture}[remember picture,overlay,shift={(current page.center)}]
\node[inner sep=0pt] at (0,0) {\includegraphics[width=\paperwidth,page=14]{arxiv_SI.pdf}};
\end{tikzpicture}

\pagenumbering{arabic}

\newpage
\maketitle
\noindent	
%\todo{Zu den Figures: DIese d[rfen eignetlich nicht ins Supplement, ich lasse sie aber bis yum Einreichen drin, damit man leichter Nachschauen kann.]}\\
\todo{Reminder vor Submission: Stimmen alle Zahlen? }
%\todo{delay klar}
%\todo{rolle csra: binding, abundance klar}

\noindent
%In this document, we develop the theoretical model for the interplay of ssDNA as an additional regulatory element with the long mRNA -- CsrA -- sRNA regulatory system. This model constitutes an extension of our hierarchical three-component model, which we presented in \cite{Schwarz2016}. Our goal is to theoretically support the hypothesis that ssDNA crucially affects the \textit{cea-cel}-delay, as seen in the experimental system.

\section{The biological system }\label{sec:biosys}

The biological phenomenon we wish to describe by means of a quantitative theoretical model is the regulation of ColicinE2 release in \textit{E. coli}. ColicinE2 is a bacterial toxin encoded by the gene \textit{cea}, which is part of the ColicinE2 operon located on a plasmid. This operon also contains genes for an immunity protein (\textit{cei} gene) and a lysis protein (\textit{cel} gene).
The lysis protein is part of the operon as cell lysis is the only way to release the toxin into the environment. Since lysis also means the death of the cell, the release of ColicinE2 is highly regulated. Previous studies have revealed the regulatory components controlling  ColicinE2 production and release on both the transcriptional and post-transcriptional levels:
\begin{itemize}
	\item The \textit{transcription} of the operon is regulated by the repressor \textit{LexA}, which is part of the \textit{E. coli} SOS response regulatory network~\cite{Janion2008,Shimoni2009}. Stressful events, such as DNA damage activate a SOS response system~\cite{Janion2008}, which stochastically triggers the transcription of the operon by degradation of \textit{LexA}. 
%	However, the level of \textit{LexA} in the cell is not reduced permanently during an SOS response, but in form of stochastic degradation events ~\cite{Shimoni2009}. 
	Once transcription starts, two mRNA transcripts are produced: \textit{short mRNA}, containing only the toxin and immunity protein, and \textit{long mRNA}, which contains also the lysis protein~\cite{Yang2010}.
	
	\item The \textit{post-transcriptional} regulation (see Fig. 1 and also Fig.~S\ref{fig:biochemNetwork}) acts on the long mRNA only. To our knowledge, its only regulator is the protein \textit{CsrA}, which binds to the Shine-Dalgarno sequence that is located on the long mRNA between the sequences coding for the immunity and lysis proteins~\cite{Yang2010}. When a CsrA protein binds to long mRNA and thus forms a complex with it, the gene for the lysis protein can no longer be translated, and thus the cell does not lyse~\cite{Yang2010}. By preventing lysis protein expression, CsrA regulates the release of ColicinE2. 
%	More specifically, free (i.e. unbound) long mRNAs, which eventually are translated to lysis proteins, can only exist if no free CsrA is left. 
	The abundance of CsrA itself is known to be regulated by the two CsrA-sequestering short RNAs (sRNAs) \textit{CsrB} and \textit{CsrC}, which both have several CsrA binding sites~\cite{Timmermans2010,Weilbacher2003}. Our current study suggests that, in addition, rings of single-stranded DNA (ssDNA) also sequester CsrA, and therefore represent a novel CsrA regulator. This ssDNA is created as an intermediate during the rolling circle replication of the ColicinE2 plasmids.
\end{itemize}

\noindent
A particular example for the importance of these regulatory interactions is the delay between production and release of ColicinE2, which has recently been studied experimentally~\cite{Mader2015}:
The translation of lysis proteins from long mRNA (and therefore lysis itself) can only start if there are free long mRNAs, that is, long mRNAs that are not bound to CsrA. 
From what is known about CsrA interactions, we assume in our biochemical model that a CsrA molecule can no longer regulate long mRNA when it is either sequestered, or degraded. %\todo{Hier noch detaillierter dass lange im Komplex gebundenes CsrA vermutlich nicht mehr bindet wenn der Komplex dissoziiert?} 
%Due to the important role of CsrA in growing bacteria,
Previous studies~\cite{Gudapaty2001a} show that CsrA is highly abundant during growth phase, mainly in form of CsrA complexes.
%Free (i.e. unbound) long mRNAs can only exist if no free CsrA proteins are present in the system to regulate them. Consequently, the translation of long mRNA to lysis protein (and therefore lysis itself) can only start if all CsrAs are either degraded or bound in complexes. 
%In growing bacteria, the production and degradation/binding rates are tuned such that CsrA is abundant. 
When an SOS response is triggered, however, the production of long mRNA increases such that free CsrA abundance decreases, and eventually lysis proteins are translated from free long mRNA.
This process does not happen instantaneously:
Due to stochasticity in the SOS response system and the time it takes to produce, bind or degrade the regulatory components involved, we find a delay between the expression of the unregulated \textit{cea} gene (part of the short mRNA) and the CsrA-regulated \textit{cel} gene (part of the long mRNA).
%Its duration is determined by the rates of production, degradation and binding for CsrA and its regulators. 
%The types and abundances of the CsrA regulators is strain dependent. In our experiments, we considered three different strains (see Table~\ref{tab:strains})
This delay is presumably not just a byproduct of regulation, but has also a biological function: It gives the cell time to accumulate ColicinE2, and thus allows for higher toxin concentrations upon the release. Moreover, it presumably also acts as a safety buffer, which prevents premature cell lysis, for instance due to fluctuations in the system~\cite{Mader2015}.

%Another aspect, which makes the time-delay particularly interesting for us, is the fact that it can be accessed experimentally.
In contrast to the abundances of the regulatory components, the \textit{cea-cel-}delay is a quantity that can readily be measured experimentally by \textit{reporter plasmids} inserted in the \textit{E. coli} cells. These reporter plasmids carry the ColicinE2 operon, but the toxin (\textit{cea}) and lysis (\textit{cel}) gene are replaced by two different fluorescence protein genes (CFP and YFP). 
Since only these two genes are replaced, the reporter plasmids have the same promoter as the ColicinE2 plasmid, and the long mRNA transcript of the reporter plasmid has the same CsrA binding site as the original long mRNA. Consequently, a reporter plasmid behaves like the ColicinE2 plasmid, but produces two types of fluorescence proteins instead of toxin and lysis proteins.
Therefore, upon introducing reporter plasmids to an \textit{E. coli} cell, one can measure the time points of production of the corresponding fluorescence proteins; these measured time points then coincide with the time points of toxin and lysis protein production. In this study, we use two reporter plasmid types, which differ by their mean abundance in the cell: type 1 (pMO3) accumulates to about 55 plasmids per cell, whereas type 2 (pMO8) only to about 13 plasmids per cell.
%The production of the fluorescent proteins then coincides with the production of the toxin or lysis protein, respectively, and thus enables us to measure the time-delay between these events.

With three different plasmid types, the original and the two reporter plasmids, we can construct five different strains (see also Table~S\ref{tab:strains}): First, the wild-type strain, C$_\text{WT}$, which carries only ColicinE2 plasmids. Inserting reporter plasmids to this strain creates, depending on the reporter plasmid type inserted, either a strain called C$_\text{REP1}$ or a strain called C$_\text{REP2}$. Completely replacing the ColE2 plasmid with one of the reporter plasmid types creates another two strains, referred to as S$_\text{REP1}$ and S$_\text{REP2}$.
%\todo{im folgenden satz gehe ich bewusst nicht naeher auf die unterschiede ein, weil da auch nicht so viel bekannt ist}
Our experiments show that the plasmid types also differ in the production of ssDNA: cells carrying the ColicinE2 plasmid do accumulate ssDNA (see Fig.~3), while this is not the case for the S$_\text{REP1}$ and S$_\text{REP2}$ strain cells, which only contain reporter plasmids (see Fig.~3 and Supplementary Information).
%The lack of ColicinE2 plasmids in the S$_\text{REP1}$ strain means that this strain does not accumulate ssDNA. 
%Due to a slightly different replication mechanism compared to the original ColicinE2 plasmid (pColE2-P9), the reporter plasmid does not accumulate ssDNA (see \todo{experimental supporting}).

%\todo{diesen absatz noch besprechen und anpassen,}
%Note that only the C$_\text{REP1}$ and the S$_\text{REP1}$ strain contain reporter plasmids. Investigating the wild-type strain using fluorescence proteins like the other two strains would require more severe genetic modifications than introducing additional plasmids. Due to the complexity of the regulatory network controlling ColicinE2 expression, this step would significantly alter ColicinE2 expression dynamics, and thus violate our results.
%Therefore, we can study the the ColicinE2 expression dynamics of both the \textit{cea} and \textit{cel} gene in the C$_\text{WT}$ only using the theoretical model presented in section~\ref{sec:mathmod}. However, it is possible to measure the lysis time of C$_\text{WT}$ bacteria (see Fig. S8), which we will use to validate our model for this strain. 
%The relevant genetic elements of these three strains presented here (C$_\text{REP1}$, C$_\text{WT}$, and S) are summarised in Table~S\ref{tab:strains}.
As discussed in detail in section~\ref{sec:mathmod}, we will use mathematical modelling to infer the delay of the wild-type strain from the delay measured in the other strains containing the reporter plasmid. Before doing so, we recapitulate the main experimental findings on the delay-times in the bacterial strains containing the reporter plasmid.

\begin{table}
	\centering
	\begin{tabular}{c c c c}
		 & number of & number of &  \\
		strain & reporter plasmids & pColE2P9 plasmids & ssDNA \\ \hline 
		S$_\text{REP1}$              & $\approx 55 $ & -- & -- \\
 		S$_\text{REP2}$              & $\approx 13$ & -- & -- \\ 
		C$_\text{REP1}$ & $\approx 55$ &  $ \approx 20$ &  accumulates \\
                C$_\text{REP2}$ & $\approx 13$ &  $ \approx 20$ &  accumulates \\ 
		C$_\text{WT}$ & -- & $\approx 20$ & accumulates \\ 
	\end{tabular} 
	\caption{The five different strains and the abundance of the genetic elements that differentiate them~\cite{Cascales2007}.}
	\label{tab:strains}
\end{table}

In our experiments (see main text) we found that the C$_\text{REP1}$ strain shows no significant delay between \textit{cea} and \textit{cel} gene expression, whereas the S$_\text{REP1}$ strain has a significant mean delay of 75 minutes. Moreover, we observed that the S$_\text{REP1}$ strain has a broad delay-time distribution around this mean value. %, whereas the delay times in the C$_\text{REP1}$ strain are all close to zero.
This raised the question as to the source of this difference. The C$_\text{REP1}$ and S$_\text{REP1}$ strains differ only in their plasmid composition and are both genetically identical (see Table~S\ref{tab:strains}).
%\todo{\@Erwin: Die beiden Plasmide enthalten tatsaechlich keine unterschiedlichen regulatorischen Elemente! Sie unterscheiden sich in der Hinsicht nur in den cea und cel Genen. Habe hier einen Satz eingefuegt, der das deutlich macht}
From this we conclude that the presence of ColicinE2 plasmids in the C$_\text{REP1}$ strain introduces further regulatory elements (compared to the reporter plasmids), which are responsible for the shorter \textit{cea-cel} delay times compared to the S$_\text{REP1}$ strain. The reporter and the ColicinE2 plasmids contain the same regulatory sequences (see \textbf{Methods}), which means that the additional regulatory elements cannot be different mRNA transcripts specifically produced by the ColicinE2 plasmid.
From what is known about the two plasmids and the regulatory network of ColicinE2 (see above), two mechanisms could in principle account for the shorter delay times in C$_\text{REP1}$, which are: First, additional production of CsrA sequestering long mRNA due to the larger plasmid copy number, and second, the accumulation of ssDNA which, as our study shows, can also sequester CsrA.

%Since the time-delay is determined by the rates of production, degradation and binding of CsrA and its regulators, this difference must originate in the specific composition of regulatory components. 
%The three investigated strains only differ in their plasmids. This lead to the hypothesis that the ssDNA, being an intermediate of the pColE2P9 replication, is a crucial regulatory component, and that the copy numbers and types of plasmids have a significant influence on the cea-cel delay.

%In the following, we develop a mathematical model for the regulation of ColicinE2 release that accounts not just for the different regulatory components present in the different strains (see Table~\ref{tab:strains}), but also the plasmid copy numbers. With this model we generate and simulate delay time distributions for the C$_\text{REP1}$ and S$_\text{REP1}$ strain, which we compare to our experimental data. Moreover, the model allows us to predict the behaviour of the C$_\text{WT}$ strain, which is inaccessible to experimental observations.

%To test and quantify the hypothesis that CsrA regulation controls the time-delay, we aim to develop a mathematical model that contains all the known components relevant for ColicinE2 regulatory network.

\section{Mathematical model of the ColicinE2 release} \label{sec:mathmod}

In the following, we develop a mathematical model that enables us to investigate the regulation of ColicinE2 release in all five strains (C$_\text{REP1}$, C$_\text{REP2}$, C$_\text{WT}$, S$_\text{REP1}$, and S$_\text{REP2}$). The model accounts for all necessary regulatory components, including the ssDNA and the different plasmid compositions.
We validate this model by reproducing the experimentally observed delay time distributions for the S$_\text{REP1}$ strain. Variation of ssDNA production in the model then allows us to quantify the impact of ssDNA production and plasmid copy number on the \textit{cea-cel} delay.
Moreover, the model enables us to infer the behaviour of the C$_\text{WT}$ strain, for which the \textit{cea-cel} delay cannot be directly measured experimentally. The inferred behaviour can be validated by comparison with experimentally measured lysis times, see Fig. S8.

%Dynamical phenomena of gene regulation have not only been studied experimentally \todo{cite}, but also using mathematical models~\cite{Legewie2008,Levine2008}.
Several experimental studies defined and probed the regulatory networks and components involved in \textit{E. coli} SOS responses, as well as ColicinE2 production and release~\cite{Yang2010,Cascales2007,Mader2015,Janion2008}. Starting from these experimental results, the regulatory interactions have also been studied using mathematical models~\cite{Shimoni2009,Schwarz2016,Ronen2002, Krishna2007}:
For the transcriptional regulation network of the \textit{E. coli} SOS response system, a stochastic model has been presented, which is able to reproduce the distribution of stochastic SOS activity peaks~\cite{Shimoni2009}. For the post-transcriptional regulation of ColicinE2 release, we recently introduced a hierarchical three-component model \cite{Schwarz2016} involving long mRNA, CsrA, and an effective sRNA. This model was also combined with the stochastic SOS signal model from Ref.~\cite{Shimoni2009} to emulate the response of ColicinE2-producing bacteria to external stress. With this combined approach, the model shows that sRNA reduces internal fluctuations and helps controlling the level of CsrA. Moreover, the model predicts stochastically distributed delays between SOS signal and lysis, which is also seen in experiments with the S$_\text{REP1}$ strain.

In this section, we extend our previous model~\cite{Schwarz2016}, taking into account the new experimental findings presented in the main text. In particular, we incorporate the additional regulator ssDNA as well as the different plasmid copy numbers and types.
%In this section, we take our three-component model~\cite{Schwarz2016} as starting point and extend it with ssDNA as additional regulatory component. 
For this step, it is important to know the derivation of the previous, three-component model, which is why we outline the derivation of the previous model as we develop our new model from scratch.
%we first give here an outline of its derivation with particular focus on the approximating methods that we will later also apply to the ssDNA. 
For a detailed derivation of the three-component model, we refer the reader to Ref.~\cite{Schwarz2016}.

\subsection{Regulatory network}\label{subsec:assumptions}

Our goal is to design a stochastic model that enables us to investigate the dynamics of the regulatory networks involved the SOS response and the ensuing synthesis and release of ColicinE2.
To this end, we first formulate the interactions of the regulatory components as a set of (deterministic) differential equations, that is, as a mass-action model. 
This approach disregards any spatial effects and considers the system as \textit{well mixed}.
%, i.e. reaction rates depend only on the total amount of molecule numbers and not on the local concentration of specific molecules.

Extending our previous study~\cite{Schwarz2016}, we build a mass-action model for the SOS response, and the regulatory network for ColicinE2 production and release from the following assumptions and properties of the components (see also Fig.~S\ref{fig:biochemNetwork}):
\begin{itemize}
	\item The abundances of long mRNA, CsrA and effective single-binding-site sRNA (see below) are denoted by $M, A$ and $S$, respectively. These abundances give the number of \textit{free} components, that is, the number of long mRNA, CsrA and sRNA molecules that are not bound in a complex. Moreover, $P_\text{COL}$ and $P_\text{REP}$ denote the copy number of ColicinE2 and the reporter plasmids, respectively (there is no need to distinguish between the two reporter plasmid types for $P_\text{REP}$ as they do not occur in the same cell at the same time).
	
	\item 
	The response to external stress (``SOS response'') is regulated by the LexA/RecA system~\cite{Janion2008}, which we incorporate into our model using the differential equations given in Ref.~\cite{Shimoni2009}. This model accounts for the production, degradation and (un)binding of the proteins LexA ($L$) and RecA ($R$), the mRNAs they are translated from ($\Ml$ and $\Mr$, respectively), as well as the number of repressed promoters controlling the transcription of these mRNAs ($\Bl$ and $\Br$, respectively).
	In this system, LexA acts as repressor: as long as a LexA protein is bound to the promoter region of the RecA or LexA operon, no mRNA is produced. 
%	Specifically, the differential equations read:
	The number of repressed RecA and LexA promoters increases if LexA binds to an unrepressed promoter, and decreases as it unbinds. Therefore, the differential equations for $\Bl$ and $\Br$ contain two terms each: a production term proportional to the abundances of LexA and unrepressed promoters, and a degradation term proportional to the number of repressed promoters. In an \textit{E. coli} cell, there is only one promoter for each LexA and RecA, which means that $\Bl$ and $\Br$ can take either the values 0 or 1. The differential equations then read
	\begin{align}
	\dot{\Br} &= \kRp(1-\Br)\Le-\kRm\Br, \label{eq:bRdot}\\
	\dot{\Bl} &= \kLp(1-\Bl)\Le-\kLm\Bl, \label{eq:bLdot}
	\end{align}
	where the $k^\pm$ denote the attachment and detachment rates of LexA to/from the promoter indicated by the subscript.	
	From unrepressed promoters the respective mRNA is transcribed, and hence, the mRNA production depends linearly on the number of unrepressed promoters. Once produced, the mRNA can spontaneously degrade. Therefore, the differential equations for the mRNAs also contain two terms each, and read:
	\begin{align}
	\dot{\Mr} &= \aMr(1-\Br)-\dMr\Mr, \label{eq:mRdot} \\
	\dot{\Ml} &= \aMl(1-\Bl)-\dMl\Ml,
	\end{align}
	where $\alpha$ and $\delta$ give the per capita production and degradation rate of the component indicated by the subscript. These mRNAs are translated to RecA and LexA proteins, respectively. Hence, the production terms of the two proteins are proportional to the respective mRNA abundance. The number of proteins decreases by spontaneous degradation. As the abundance of RecA is only affected by these two processes, its differential equation reads:
	\begin{align}
	\dot{\R} &= \bR\Mr - \dR\R, \label{eq:Redot}
	\end{align}
	where $\bR$ and $\dR$ denote the per capita production and degradation rate of RecA.
	Since LexA acts as regulator in the LexA/RecA-system, its abundance is also affected by the interactions with the promoters. Consequently, the terms from eqs.~\eqref{eq:bRdot} and~\eqref{eq:bLdot} appear in the differential equation for LexA, but with opposite sign. Moreover, in case of an SOS signal, RecA depletes LexA, motivating an additional degradation term bilinear in $\Le$ and $R$, with the degradation constant $c_p$. 
	Apart from these interactions with the LexA/RecA-system, LexA is also the repressor of the colicin operon. Therefore, the LexA/RecA SOS response system interacts with the regulatory system of ColicinE2 production and release via $B$, the number of repressed promoters of the ColicinE2 operon. Its differential equation contains two terms analogous to the LexA and RecA promoters:
	\begin{align}
	\dot{B} &= k_P^+ (P_\text{COL}+P_\text{REP}-B)\Le-k_P^- B, \label{eq:bGdot} 
	\end{align}
	where the $k^\pm_P$ denote the attachment and detachment rates of LexA repressor to/from the ColicinE2 promoter. Unlike $\Br$ and $\Bl$, $B$ can take values between 0 and $P_\text{COL}+P_\text{REP}$. The terms from eq.~\eqref{eq:bGdot} apppear, again with opposite sign, also in the differential equation for LexA, which, altogether, reads 
	\begin{align}
	\dot{\Le} &= \bLe\Ml -\dLe\Le -\kLp(1-\Bl)\Le +\kLm\Bl -\kRp(1-\Br)\Le \nonumber \\
	&\quad  +\kRm\Br -k_P^+ (P_\text{COL}+P_\text{REP}-B)\Le +k_P^- B -c_p\R\Le,  \label{eq:Ledot}
	\end{align}
	where $\bLe$ and $\dLe$ give the per capita production and degradation rate of LexA. 
	For a detailed discussion of these equations, we refer to Ref.~\cite{Shimoni2009}.
	Note that the SOS response system, eqs.~\eqref{eq:bRdot}-\eqref{eq:Ledot}, interacts with the ColicinE2 regulatory network only through the parameter $B$ (see also next bullet point).

	\item The total production rate of long mRNA in the cell is proportional to the number of unrepressed ColicinE2 promoters in the cell. This number is given by the total number of plasmids in the cell, $P_\text{COL}+P_\text{REP}$, minus $B$, the number of promoters with the repressor LexA bound to it. Hence, the production rate of long mRNA reads
	\begin{align} \label{eq:alphaM}
	\alpha_M (P_\text{COL}+P_\text{REP}-B),
	\end{align}
	where $\alpha_M$ is the production rate per unrepressed promoter. Note that considering different plasmid types generalizes our earlier work presented in Ref.~\cite{Schwarz2016}.
%	The abundance of free long mRNA in the cell is denoted as $M$. 
%	\item In our discussion of both the hierarchical three-component model and its extension, we always assumed an average long mRNA production rate, $\aM$. However, the long mRNA production in fact depends heavily on the state of the SOS promoter (bound = repressed, unbound = unrepressed), which itself depends on the SOS response~\cite{Yang2010}. In order to account for the effects of an SOS signal to our post-tran\-script\-ion\-al model, we describe the SOS response system by a stochastic model as suggested by Shimoni \cite{Shimoni2009}. 
%	
%	This model for the SOS response comprises the repressor of the ColicinE2 promoter, \textit{LexA}, as well as its binding dynamics to the promoter region of the plasmids. Therefore, it reproduces the number of unrepressed promoters, $B_\text{REP1}$. We can then connect the SOS response to the post-transcriptional model by replacing the average long mRNA production rate, $\aM$:
%	\begin{align} \label{eq:alphaM}
%	\alpha_M (P_\text{REP}+P_{Rep}-B_\text{REP1}),
%	\end{align}
%	where we have also included the number of reporter plasmids, $P_{Rep}$. Note that \textit{for the long mRNA production} there is no difference between the ColicinE2 and the reporter plasmid in the mathematical model, as both of them contain the same regulatory elements and produce equally regulated mRNA transcripts. The fact that only the ColicinE2 plasmid produces ssDNA (due to rolling circle replication) is only relevant for the ssDNA production.

	\item CsrA is produced at a constant rate, $\alpha_A$. 
%	The abundance of free CsrA in the cell is denoted as $A$. 

 	\item The ColicinE2 system has two different regulatory sRNAs: CsrB and CsrC. Apart from having different numbers of CsrA binding sites and slightly different half-lifes, their mode of binding with CsrA is very similar. Hence, we assume that we can describe their regulatory impact by a single \textit{effective sRNA} with corresponding effective parameters (see Ref.~\cite{Schwarz2016} for details).
 	Using effective sRNAs in a mathematical model is indeed supported by experiments, which show that the knock-out of either CsrB or CsrC causes a compensating overproduction of the other sRNA (see the main text, and Ref.~\cite{Weilbacher2003}). This compensation is a natural consequence of a positive regulatory effect of CsrA abundance to sRNA production (see bullet point below), and highlights the functional equivalence of CsrB and CsrC. 
 	In Ref.~\cite{Schwarz2016} we also showed that this effective sRNA, which contains $N \approx 10$ CsrA binding sites is equivalent to $N$ \textit{effective single-binding-site sRNAs}. This drastically reduces the mathematical complexity of the model.
	
	\item Several studies found that the production of CsrB and CsrC is indirectly regulated by the abundance of CsrA via the BarA/UvrY-system~\cite{Yang2010,Timmermans2010,Weilbacher2003}. Since the details of this interaction are largely unknown, we model this positive regulation with an sRNA production rate that is a linear function of the CsrA abundance. In addition to this linear term, we also introduce a constant baseline production term, since studies show that sRNAs are also produced (at very low levels) in the absence of free CsrA~\cite{Weilbacher2003}. Both production terms contain a factor $N$, as  we consider effective single-binding-site sRNAs in our model (see the previous bullet point). The production term of the effective single-binding-site sRNAs thus reads
	\begin{align*}
	\alpha_{S,0} N + \alpha_{S,c} A \cdot N,
	\end{align*}
	with the baseline production rate $\alpha_{S,0}$, the linear coupling coefficient $\alpha_{S,c}$, and the abundance of CsrA proteins $A$. 
%	The abundance of free effective single-binding-site sRNAs in the cell is denoted as $S$. 
	Note that we did not consider the positive feedback of CsrA on sRNA production in Ref.~\cite{Schwarz2016}. 
%	We included this feedback in our new model, as it showed to stabilize the system towards parameter changes.
	
	\item The degradation rates of long mRNA, CsrA and the effective sRNA are each proportional to their respective abundance, and read $\dM M$, $\dA A$ and $\dS S$, respectively.
	
	\item CsrA can bind to both long mRNA and the effective sRNA, and thus forms CsrA-long mRNA and CsrA-sRNA complexes ($C_{MA}$ and $C_{SA}$, respectively). In line with previous studies~\cite{Legewie2008,Levine2008} and our three-component model~\cite{Schwarz2016}, we assume that the formation and disassembly of these complexes is much faster than the other processes involved in post-transcriptional regulation. Therefore, we can employ adiabatic elimination, $\partial_t C_{MA} \equiv 0$ and $\partial_t C_{SA} \equiv 0$.
	In Ref.~\cite{Schwarz2016}, we show that this enables us to combine the formation, disassembly and degradation of the complexes into effective binding parameters, $\kM$ and $\kS$. As a consequence, we can solve for the complex abundances, $C_{MA}$ and $C_{SA}$, and eliminate them from our set of differential equations (see Ref.~\cite{Schwarz2016} for details).
%	The CsrA-long-mRNA-complex cannot be translated to lysis proteins.
%	This allows us to eliminate the differential equations for the complexes from our mass action model.
%	use adiabatic elimination of fast processes to

 	\item The precise mechanism for the degradation of CsrA-sRNA and CsrA-long mRNA complexes is not known. Here, we assume that\textit{ CsrA dimers are always degraded once their complex partner is degraded} (in other words: CsrA cannot ``survive'' the degradation of its partner). 
% 	\todo{\@Erwin: wir koennen den anderen fall simulieren. wenn man die annahme lockert, wird weniger CsrA abgebaut. ich wuerde den effekt aber gering erwarten}

	\item CsrA is a main regulator in growing \textit{E. coli} cells, which is known to bind to over 700 different targets~\cite{Edwards2011, Yakhnin2012}. In the Supplementary Information of Ref.~\cite{Schwarz2016} we show how one can eliminate the many targets of CsrA to obtain a reduced system, which contains only the components that are changed by the processes the model focusses on (in this case: SOS-induced production and release of ColicinE2). In the mathematical model presented in this section, we reduce the system to three CsrA targets: long mRNA, sRNA, and (see below) ssDNA. In agreement with experiments~\cite{Gudapaty2001a}, the production rate of the effective sRNA is large compared to the production of long mRNA and ssDNA, such that the vast majority of sequestered CsrA proteins is bound to sRNA.

%	, which are also shown in Fig. S4, cannot be reproduced by our model. This is because our model does not account for the many other regulatory networks involving CsrA, whose effects are subordinate in the presence of the two sRNAs\todo{CITE quellen von Alex}, and only become important in the special case of the double-sRNA deletion mutant. 

	\item The short mRNA is not regulated by CsrA, and hence not part of the regulatory network. However, our experiments use the translation of short mRNA (specifically, the translation of the \textit{cea} gene) as proxy for promoter activity in the S$_\text{REP1}$, S$_\text{REP2}$, C$_\text{REP1}$, and C$_\text{REP2}$ strain. To enable the experimental validation of our model, we include the production of short mRNA in our model. Due to the lack of regulation, the corresponding differential equation is  decoupled from $M, A$ and $S$, and reads (with $\frac{\partial M_\text{short}}{\partial t} \equiv \dot{M_\text{short}}$)
	\begin{align}
	 \partial_t M_\text{short} = \alpha_{M_\text{short}} - \delta_{M_\text{short}} M_\text{short},
	\end{align}
	where $\alpha_{M_\text{short}}$ and $\delta_{M_\text{short}}$ are the rate constants for production and per-capita degradation, respectively.
%	
% 	\item We \textit{do not model the toxin and lysis proteins}, as they are not directly regulated by the post-transcriptional system we discuss. 
\end{itemize}
The properties and assumptions of the SOS response and the ColicinE2 regulatory system we listed above have already been used (if not stated otherwise) in the combined model for SOS response and ColicinE2 regulation presented in Ref.~\cite{Schwarz2016}.

\paragraph*{ssDNA as regulatory component:}
In the main text, we show experimentally that single-stranded DNA (ssDNA) serves as a component of post-transcriptional regulation of ColicinE2 production and release.
Since this is a novel and, so far, an undocumented role of ssDNA, we briefly discuss how it acts as a regulator for CsrA in an \textit{E. coli} cell. 

ssDNA is an intermediate in the rolling circle replication mechanism of the ColicinE2 plasmid: The plasmid consists of double-stranded DNA (dsDNA). The first step in its replication is the production of a ring-shaped ssDNA transcript. These transcripts are produced both in absence and presence of an SOS signal (Fig. S5), which means that ssDNA production is constant. It is assumed that once a ring of ssDNA is completed, it detaches from the plasmid and diffuses freely through the cell. During this time, it is converted to double-stranded DNA, which eventually results in a new plasmid. Between the detachment of the single-stranded ring and the formation of a new plasmid, the ssDNA acts as a regulator of CsrA: Since the ssDNA includes the coding sequences present in the long mRNA, CsrA can bind to the Shine-Dalgarno sequence of the \textit{cel} gene located on the ssDNA, and thus forms an ssDNA-CsrA complex. This allows the ssDNA to regulate free CsrA levels by sequestration, similar to the CsrA regulation by sRNA. 
%In the following, we give a detailed description on how we include this additional regulatory component to our model.

%Introducing ssDNA to the three-component model means that we have to account for the following additional properties and assumptions:
In our mathematical model, we account for these properties of ssDNA as follows:
\begin{itemize}
	\item The production rate of ssDNA is assumed to be proportional to the number of ColicinE2 plasmids, $P_\text{COL}$, as it is an intermediate product of the rolling circle replication mechanism of the ColicinE2 plasmid. It reads
	\begin{align}
		\alpha_D \cdot P_\text{COL},
	\end{align}
	with the per plasmid production rate constant $\alpha_D$.
	
	\item The degradation of ssDNA is proportional to the ssDNA abundance, $D$, and thus reads  
	\begin{align}
		\delta_D \cdot D,
	\end{align}
	with the per capita degradation rate constant is  $\delta_D$.
	
	\item The ssDNA has two binding sites for a CsrA dimer (see \textbf{Methods} of the main text), and thus can form a complex, $C_{DA}$, with it. Complex formation occurs with rate $k_D^+$, and the complexes dissociate into ssDNA and a CsrA dimer with rate $k_D^-$. Apart from disassembly, we also include the possibility that a CsrA-ssDNA-complex can spontaneously degrade (meaning that both CsrA and ssDNA are degraded at the same time) by introducing the per capita rate $\delta_{DA}$. The abundance of complexes is denoted by $C_{DA}$. 

\end{itemize}
Taken together, we can now formulate a set of differential equations, which allows us to quantify these interactions.
These interactions are also illustrated as a biochemical network in Fig.~S\ref{fig:biochemNetwork}.

We begin with the differential equation for the time evolution of the long mRNA, $M$.
From the properties collected above, we conclude that this equation must contain three terms: The first term describes the production of long mRNA, which is proportional to the number of unrepressed promoters. This number is calculated from the difference between the total plasmid copy number, $P_\text{COL}+P_\text{REP}$, and the number of repressed promoters, $B$.
%We refer the reader to the Supporting Information of Ref.~\cite{Shimoni2009} for a detailed discussion of the differential equations of this subsystem.
The abundance of long mRNA is reduced by a second and a third term: The second term describes the spontaneous degradation of long mRNA, and is proportional to its abundance, $M$. The third term is bi-linear (that is, it is proportional to $A$ and $M$) and represents the effective coupled degradation of long mRNA in complexes with CsrA. This term combines the binding of CsrA to long mRNA, the dissociation of this complex, and its degradation in an effective binding parameter $k_M$. The three terms  read:
\begin{align}
\dot{M} & = \alpha_M (P_\text{COL}+P_\text{REP}-B) - {\dM}M - {\kM}M \cdot A. \label{eq:3rateM}
\end{align}
The derivation of the effective coupled degradation in the third term is described in detail in Ref.~\cite{Schwarz2016}; an analogous derivation for the ssDNA is given below. The $B$ in the first term is determined by the LexA/RecA subsystem of the SOS response, in particular by eq.~\eqref{eq:bGdot}.

The differential equation for the time evolution of the effective single-binding-site sRNA, $S$, consists of terms very similar to that for long mRNA.
Two terms account for spontaneous and effective coupled degradation, respectively, and are structurally the same as in eq.~\eqref{eq:3rateM}. This is due to the fact that the sRNAs regulate CsrA in the same way as CsrA regulates the long mRNA, by forming complexes. The production term is, however, different, and contains two parts: The first part, $\alpha_{S,0}$, describes a constant baseline production, which ensures the production of sRNAs in the absence of CsrA. The second part depends linearly on the abundance of free CsrA, and thus accounts for the positive regulatory function of CsrA for the sRNAs. Taken together, these four terms give the differential equation for $S$:
\begin{align}
\dot{S} & = \alpha_{S,0} N + \alpha_{S,c} N \cdot A-{\dS}S-\kS S\cdot A.\label{eq:3rateS}
\end{align}

Having described the two partners of CsrA, we now turn to the differential equation for CsrA itself. Again, this equation has a very similar structure to eqs.~\eqref{eq:3rateM} and~\eqref{eq:3rateS}: An, in this case constant, production term, as well as a term for spontaneous degradation. Here, however, we have more than one coupled degradation term, since CsrA can bind to more than one component: long mRNA ($M$), sRNAs ($S$), and ssDNA ($D$). The effective coupled degradation terms for long mRNA and sRNA are exactly the same as in eqs.~\eqref{eq:3rateM} and~\eqref{eq:3rateS}, respectively. This reflects the fact that the formation of a long-mRNA/CsrA- or sRNA/CsrA-complex has for both complex partners the same consequence, that is, it reduces the abundance of free CsrA by 1. The coupled degradation part is also responsible for the hierarchical regulation, which we discussed in~\cite{Schwarz2016}: The actual regulation target, long mRNA ($M$), exclusively binds to CsrA; the sRNAs affects the free long mRNA level only indirectly by sequestering the CsrA and thus ``regulating the regulator''. Moreover, we also have to account for ssDNA/CsrA-complexes. Since we have not derived an effective coupled degradation for this complex yet, we explicitly account for its formation and disassembly. This means that we have to include two terms that account for the decrease of free CsrA due to the formation of ssDNA/CsrA-complexes and the increase of free CsrA when such a complex disassembles. 
Altogether, the differential equation for CsrA reads
\begin{align}
\dot{A} & = {\bA} - {\dA}A  - \kM M\cdot A - \kS A\cdot S \color{red} - k_{D}^+ D \cdot A + k_{D}^- C_{DA}, \label{eq:3rateA}
\end{align}
where $C_{DA}$ is the abundance of ssDNA/CsrA-complexes, and $ k_{D}^\pm$ the complex binding and disassembly rate, and the terms containing the novel regulator ssDNA are highlighted in red.
Note that we consider the ssDNA to have only one binding site for CsrA in our model. We account for the second binding site analogously to the many binding sites of the sRNA, that is by assuming $D$ to be an effective, single binding site ssDNA, with an effective production rate fitted to experimental data. 
%This effective production rate is close to the actual value, since our experiments show that the two binding sites are rarely occupied at the same time. %\todo{cite} 

The two ssDNA terms highlighted in red also appear in the differential equation for $D$, since the formation and disassembly of ssDNA/CsrA-complexes in- and decreases also the abundances of ssDNA. The spontaneous degradation is accounted for by a separate degradation term, already known from the differential equations of the other components.
The production term of ssDNA is proportional to $ P_\text{COL}$, the number of ColicinE2 plasmids in the cell, since ssDNA is an intermediate of the ColicinE2 plasmid replication. The differential equation for ssDNA therefore reads
\begin{align}
\dot{D} &  = \alpha_D P_\text{COL} -\delta_D D - k_{D}^+ D\cdot A + k_{D}^- C_{DA}. \label{eq:3rateD} 
\end{align}

We are still left with the dynamics of the ssDNA-CsrA-complexes, $C_{DA}$.
The ``production'' term of the ssDNA/CsrA-complexes is the binding term already known from eqs.~\eqref{eq:3rateD} and~\eqref{eq:3rateA}, but in this case with a positive sign. The number of complexes is reduced by complex disassembly, which is accounted for by the term $k_{D}^- C_{DA}$ that also appears in eqs.~\eqref{eq:3rateD} and~\eqref{eq:3rateA} with a different sign. Apart from complex disassembly, the complexes can be degraded (in the sense that the complexes and their components are destroyed) spontaneously, which is given by a spontaneous degradation term. Taken together, the differential equation for ssDNA reads
\begin{align}
\dot C_{DA} & = k_{D}^+ D\cdot A - k_{D}^- C_{DA} -\delta_{C_{DA}} C_{DA}. \label{eq:3rateCDA}
\end{align}

\paragraph{Effective coupled degradation of ssDNA and CsrA:}
In the discussion of the ssDNA properties, we saw that ssDNA also has a Shine-Dalgarno sequence, just as the long mRNA. This suggests that we can make the same assumptions for the CsrA-ssDNA-complex as we did for the CsrA-long-mRNA-complex.
In the following, we proceed analogously to the simplification of the hierarchical three component model (see Ref.~\cite{Schwarz2016}), and assume fast dynamics of complexes. Adiabatic elimination  ($\dot C_{DA} \equiv 0$) yields
\begin{align} \label{eq:cda}
C_{DA} = \frac{k_D^+ DA}{k_D^- + \delta_{C_{DA}}} =\frac{k_{D} DA}{\delta_{C_{DA}}},
\end{align}
with the effective binding parameter 
\begin{align*}
k_D := \frac{k_D^+\delta_{C_{DA}}}{k_D^-+\delta_{C_{DA}}}.
\end{align*}
By inserting eq.~\eqref{eq:cda} into eqs.~\eqref{eq:3rateA}-\eqref{eq:3rateCDA}, we get our final set of differential equations, which includes all four components:
\begin{align}
\dot{M} & = \alpha_M (P_\text{COL}+P_\text{REP}-B) - {\dM}M - {\kM}M \cdot A, \label{eq:rateM}\\
\dot{S} & = \alpha_{S,0} N + \alpha_{S,c} N \cdot A-{\dS}S-\kS S\cdot A, \label{eq:rateS}\\
\dot{A} & = {\bA} - {\dA}A  - \kM M \cdot A - \kS S\cdot A -k_D D \cdot A \label{eq:rateA}\\
\dot{D} & = \alpha_D P_\text{COL} - \delta_D D - k_D D \cdot A \label{eq:rateD}.
\end{align}

The new regulative component ssDNA acts in the same fashion as the sRNA by binding CsrA. Compared to the original three component system, eqs.~\eqref{eq:3rateM}-\eqref{eq:3rateA}, the extension with ssDNA therefore resulted in a system of equations with the same types of terms (source term, spontaneous degradation, coupled degradation). 
We use eqs.~\eqref{eq:rateM}-\eqref{eq:rateD} to study gene expression dynamics for all three different strains. This is done by adjusting the corresponding values for $P_\text{COL}$ and $P_\text{REP}$, see section~\ref{sec:paramvalues}. Moreover, we investigate the impact of ssDNA on the regulation of ColicinE2 production and release.

\section{Parameter values} \label{sec:paramvalues}

For the parameters associated with long mRNA, CsrA and the effective sRNA, we adjusted the values that we determined in our previous study (\cite{Schwarz2016}) according to new measurements. In particular, they  were chosen such that they are in accordance with our own experimental measurements ($\kM$ and $\kS$) or other studies (see below). In particular, the rates read (given per \textit{E. coli} cell volume, and using the shorthand notation ``\#'' for molecule numbers):
\begin{table*}[h]
	\centering
	\begin{tabular}{c|c|c|l}
		rate const. & value & unit &   description\\ \hline 
		$\alpha_M$ & 0.05 & min$^{-1}$ & production of long mRNA \\ 
		$\alpha_{S,0} N $& 0.1 & min$^{-1}$& baseline production of eff. sRNA\\ 
		$\alpha_{S,c} N $& 0.07 & min$^{-1}\cdot $\#$^{-1}$ & production factor of eff. sRNA\\ 
		$\alpha_A$ & 4.5	& min$^{-1}$ & production of CsrA \\ 
		$\dM$ & 0.04 	& min$^{-1}\cdot $\#$^{-1}$ &  degradation of long mRNA \\ 
		$\dS$	&  0.023 & min$^{-1}\cdot $\#$^{-1}$ & degradation of effective sRNA\\ 
		$\dA$	&   0.00007 & min$^{-1}\cdot $\#$^{-1}$  & degradation of CsrA\\ 
		$\kM$	&   0.007 & min$^{-1}\cdot $\#$^{-2}$  & eff. binding of CsrA to long mRNA \\ 
		$\kS$	&  0.011 & min$^{-1}\cdot $\#$^{-2}$  & eff. binding of CsrA to sRNA
	\end{tabular} 
\end{table*}

\noindent
The three degradation rates ($\dM,\dS,\dA$) were determined in previous, experimental studies~\cite{Gudapaty2001a,Weilbacher2003}. The production rates ($\alpha_A, \alpha_{S,c}, \alpha_M$) were fitted such that they reproduce component abundances from experimental studies~\cite{Gudapaty2001a}. The baseline production for the sRNAs, $\alpha_{S,0} N $, is set to a low value, as only few sRNAs are produced in the absence of CsrA~\cite{Weilbacher2003}.

In Ref.~\cite{Schwarz2016} we showed that a Poisson-distributed plasmid copy number gives very similar results to a fixed plasmid copy number. This is due to the fact that plasmid replication happens on larger timescales than the regulatory interactions considered in our model. We retain this simplifying assumption, and set the number of ColicinE2 plasmids constant at $P_\text{COL}=20$, which is the average value~\cite{Cascales2007}. For the reporter plasmids in the C$_\text{REP1}$, C$_\text{REP2}$, S$_\text{REP1}$ and S$_\text{REP2}$ strains, we take for type 1 the average copy number $P_\text{REP}=55$~\cite{Megerle2008}, and for type 2 the average copy number $P_\text{REP}=13$, which we both also assume constant. 

Adding ssDNA dynamics to the system introduces three new effective rates, $\alpha_D, \delta_D,$ and $k_D$. 
%As the precise values for these quantities are hard to obtain, we give estimates motivated from similar biological situations.
% As the CsrA binding site on the ssDNA is the same as on the long mRNA (Shine-Dalgarno sequence), 
We assume that the ssDNA and the mRNA are equally stable, and therefore use the same degradation rate constants for both:
\begin{align*}
\delta_D \equiv \delta_M &= 0.04 \text{ min}^{-1}\cdot \text{\#}^{-1}.
\end{align*} 
In combination with the K$_\text{D}$-value measurements for ssDNA we could determine the coupled degradation constant to
\begin{align*}
k_D &= 0.0001 \text{ min}^{-1}\cdot \text{\#}^{-2}.
\end{align*} 
Finally, we have to define the value for the production rate constant of ssDNA, $\alpha_D$, which has not been explicitly measured yet. 
%Due to the larger size of ssDNA compared to long mRNA, we assume this rate to be smaller than the production rate constant of long mRNA, and therefore use values in the range between 0 and $\alpha_M$:
% Due to the similar role of ssDNA and long mRNA, we assume this rate to be of similar order as the production rate constant of long mRNA:
However, our experimental data suggests that ssDNA accumulates abundances about an order of magnitude larger than long mRNA. From fitting the ssDNA production to this rough abundance relation, and also to measured delay-times, we obtain
\begin{align*}
\alpha_D = 7 \text{ min}^{-1}\cdot \text{plasmid}^{-1}.
\end{align*}
To study the influence of ssDNA on the \textit{cea-cel} delay, we varied the value of $\alpha_D$ between 0 and 9, see Fig.~S\ref{fig:hist_alphaD} and~S\ref{fig:dyn_alphaD}. For the validation of our model, we also tested various values of $\alpha_{S,c}$ and $\alpha_M$ (data not shown). These tests showed that, in general, the model is robust to parameter variations, in the sense that changing a parameter value by a few percent only had minor consequences for the resulting delay times and component abundances. 
%\todo{sollen wir hier noch erwaehnen das im falle CsrA produktionsrate das feedback zu sRNA verantwortlich dafuer ist?}

\section{Simulation results}\label{sec:results}
                                     
The differential equations eqs.~\eqref{eq:rateM}-\eqref{eq:rateD} give, in combination with the SOS response model, eqs.~\eqref{eq:bRdot}-\eqref{eq:Ledot} (see~\cite{Shimoni2009}), a description of the regulatory interactions governing ColicinE2 production.
They enable us to study steady states and the deterministic dynamics of gene expression for all five strains.
However, the SOS response~\cite{Shimoni2009} shows an inherent stochasticity: The ColicinE2 promoter is not activated permanently during an SOS signal, but in stochastically appearing bursts of activity. Moreover, most of the regulatory components like long mRNA occur in low abundances, such that also intrinsic demographic fluctuations  in the ColicinE2 regulatory system become important.
We can study stochastic effects like these by formulating the deterministic dynamics described in eqs.~\eqref{eq:bRdot}-\eqref{eq:Ledot} and  eqs.~\eqref{eq:rateM}-\eqref{eq:rateD} as a stochastic process. To this end, we consider each component ($M, S, A$ and $D$, as well as the components of the SOS response system) as random variables that are changed by stochastic events like production or degradation of molecules. Each of these events occurs at an average rate that equals the corresponding term in the mass action model. For instance, the effective coupled degradation of a long mRNA and CsrA happens at a rate $k_M M \cdot A$ (see eq.~\eqref{eq:rateM}), which decreases both the abundance of long mRNA ($M$) and the abundance of CsrA ($A$) by 1. By defining all remaining stochastic production, degradation and binding events in the system this way, we obtain a description of the SOS response and ColicinE2 regulatory system as a stochastic (Markov) process. We then use the Gillespie algorithm~\cite{Gillespie1977} to implement the stochastic process as a stochastic simulation. This simulation enables us to produce stochastically correct realisations of the temporal evolution of the system's random variables.
%enable us to study the three different strains (by setting the corresponding values for $P_\text{COL}$ and $P_\text{REP}$, see section~\ref{sec:paramvalues}), as well as the impact of ssDNA on the regulation of ColicinE2 production and release.
%To validate the theory by experiments, these stochastic simulations were implemented such that they give the delay times between \textit{cea} and \textit{cel} expression.
The results from sufficiently large ensembles of these realisations is then the basis for the validation of the theory by experimental data.

In our simulations, we followed the scheme already developed in Ref.~\cite{Schwarz2016}: We initiate the system in a non-SOS state, where the parameter $c_p$ in eq.~\eqref{eq:bGdot} of the SOS response system (see also Ref.~\cite{Shimoni2009}) is set to 0, that is, RecA does not cleave LexA. Therefore, $B$, the number of unrepressed promoters, is low (1 for S$_\text{REP2}$, 2 for C$_\text{WT}$, 3 for C$_\text{REP2}$, 4 for S$_\text{REP1}$, and 5 for C$_\text{REP1}$).
After 200 minutes, we mimic the effect of an SOS signal by increasing the parameter $c_p$, such that RecA catalyses the degradation of LexA, which acts as repressor for the ColicinE2 operon. This has the effect that the production of both long and short mRNA immediately increases. The SOS signal is stopped again at $t=500$ minutes. 
%We define the first point in time at which more than 5 free long mRNAs exist in the system as ``\textit{cel} expression''. 
For each set of parameters, this scheme is repeated \realisations times in order to obtain an ensemble of \realisations realisations. 

To be able to compare these simulation results with experiments, we have to give an appropriate definition of the \textit{cea-cel} delay in the simulations.
%Our simulations show that the CsrA abundance decreases during the SOS signal, as more CsrA-sequestering long mRNA is produced. Once there is no free CsrA left, we find free long mRNA in the system. We define the first point in time at which more than \cp free long mRNAs exist in the system as ``\textit{cel} expression'', since already a few mRNAs suffice to produce enough lysis proteins to initiate cell lysis \todo{cite}.
%As the beginning of ``\textit{cea} expression'', we define the point in time at which the short mRNA level rises to two times \todo{final checken} its value before the SOS signal started. 
As the beginning of ``\textit{cea} expression'', we define the point in time at which the short mRNA level rises to two times its value before the SOS signal started. 
Our simulations show that the CsrA abundance decreases during the SOS signal, as more CsrA-sequestering long mRNA is produced. Once there is no free CsrA left, we find free long mRNA in the system. We define the first point in time at which more than \cp free long mRNAs exist in the system as ``\textit{cel} expression''. This definition accounts for the fact that in general fewer lysis proteins are produced than toxin proteins~\cite{Cascales2007}.
In the experiment, the expression of \textit{cea} and \textit{cel} are defined by the point in time the respective fluorescence intensity reaches five times its basal (i.e. pre-SOS) level. Therefore, the expression times are determined by the appearance of proteins in the experiment, but by the appearance of mRNAs in the stochastic simulations. 
We choose the different definition of the delay in the simulations, as the specific biochemical rates of many processes involving the mRNAs and proteins are largely unknown, and have to be fitted according to observed abundances. If we included the translation of short and long mRNA to toxin, lysis and fluorescence proteins used in the experimental study into our model as well, we would add several new parameters that require fitting to our mathematical model, without getting a more precise definition of the thresholds that determine the delay.
%We did not include the translation of mRNAs to our model, since this step makes the model more complex without avoiding the necessity to define \textit{cea} and \textit{cel} expression thresholds.
%The lower thresholds in the model (factor 2 vs. factor 5 in case of the \textit{cea} expression) is motivated by the fact that several proteins are translated from a single mRNA (translational burst). 
Moreover, comparing a delay in the production of mRNAs with a delay in the production of proteins is valid in our case, since both fluorescence proteins have very similar maturation times~\cite{Mader2015}, and since we are interested in the relative rather than the absolute times of protein expression.

For the C$_\text{WT}$ strain, we cannot compare the \textit{cea-cel} delay-time from our simulations with experimental results due to the lack of reporter plasmids (see main text). To still be able to validate our stochastic simulations with experimental data in this case, we use the time between the beginning of the SOS response and cell lysis (referred to as ``lysis time''), which we also measured in experiments (see Fig.~S8). The absolute values of the lysis time will differ between our simulations and experiments, as the simulations do not account for maturation times and other processes of equal duration in all three strains. Therefore, we do not compare lysis times themselves, but the differences of the C$_\text{REP1}$ and C$_\text{WT}$ strain's lysis times, which eliminates these constant factors.

In the following, we discuss the results of our stochastic simulations for the  experiments presented in the main text.

\subsection{The role of CsrA}

As a first step to validate the mathematical model for the S$_\text{REP1}$ strain (that is, with no ssDNA in the system), we test the role of CsrA for the \textit{cea-cel} delay.
The corresponding experiment varied the binding affinity of CsrA to long mRNA, see Fig.~2E in the main text. Specifically, the experiment measured the average \textit{cea-cel} delay-time for the original S$_\text{REP1}$ strain, and for two mutant strains (CsrA1 and CsrA2) with higher and lower CsrA binding affinity, respectively. The results of these experiments (see Fig.~2E) showed that an increased CsrA binding affinity leads to longer average \textit{cea-cel} delay-times, since the regulation of long mRNA by CsrA sequestration happens more effectively with stronger CsrA binding. Consequently, we found a much shorter average \textit{cea-cel} delay for the mutant with lower binding affinity.
In our mathematical model defined by eqs.~\eqref{eq:rateM}-\eqref{eq:rateD},
the binding affinity of CsrA for long mRNA relates to the parameter $\kM^+$, which appears in the effective coupled degradation parameter, 
$\kM = \frac{\kM^+\delta_{{\cMA}}}{\kM^-+\delta_{{\cMA}}}$ 
(see Ref.~\cite{Schwarz2016} for details on this equation). 
% Therefore, we can emulate higher and lower affinities of CsrA for the long RNA by raising or reducing the $\kM$ value.
However, the experiments do not measure the rates $\kM^+$ and $\kM^-$, but their ratio $\kM^-/\kM^+$, known as K$_D$ value (see Fig. 2D of the main text). Since the degradation rate of the complexes is much lower than the dissociation rate (this follows from the fast complex equilibration assumption), we can determine $\kM$ from the complex degradation and the K$_D$ values:
\begin{align} \label{eq:kMfromKD}
\kM = \frac{\kM^+\delta_{{\cMA}}}{\kM^-+\delta_{{\cMA}}}
    \approx \frac{\kM^+\delta_{{\cMA}}}{\kM^-}
    = \frac{\delta_{{\cMA}}}{K_D}
\end{align}
We used eq.~\eqref{eq:kMfromKD} and K$_D$ values given in Fig. 2D to determine the binding parameter constants $k_M$ for the S$_\text{REP1}$ strain and its two mutants CsrA1 and CsrA2.
Using these results, we then performed a set of numerical simulations, and recorded the \textit{cea-cel} delay-times. 
The mean delay times for different values of $k_M$ are shown in main text Fig.~2F, where we also give the parameter values.
% where we have chosen the different values of $k_M$ such that they fit the corresponding experimental data. 
For the \textit{cea-cel} delay-times we find the same behaviour as in the experiments: higher values of $k_M$ result in broader delay time distributions with a larger average delay time, whereas smaller values give a narrow delay time distribution with an average delay close to the start of the SOS signal. 
%In the absence of CsrA binding to long mRNA, $k_M=0$ (not shown), the regulation is made void as free long mRNA is produced even without an SOS signal.
Our results show that the binding of CsrA to long mRNA has a key influence on the delay time distribution. This highlights the critical role of CsrA for the delay between toxin production and release. 
Therefore, we expect regulative components affecting the abundance of CsrA to have an indirect effect on the duration of the \textit{cea-cel} delay.

\subsection{\textit{cea-cel} delay in the five different strains }
%C$_\text{REP1}$,C$_\text{REP2}$, C$_\text{WT}$, S$_\text{REP1}$ and S$_\text{REP2}$

%\todo{Einzeltrajektorien in Figure}
The experiments discussed in the main text show that the C$_\text{REP1}$,C$_\text{REP2}$, C$_\text{WT}$, S$_\text{REP1}$ and S$_\text{REP2}$ all have very different \textit{cea-cel} delay times: The C$_\text{REP1}$  strain, for instance, lyses almost immediately after \textit{cea} expression, whereas the S$_\text{REP1}$ strain shows a significant \textit{cea-cel} delay. In this section, we employ stochastic simulations of the model described in eqs.~\eqref{eq:rateM}-\eqref{eq:rateD} to study the origin of this difference and the effects of ssDNA production. Moreover, we infer the \textit{cea-cel} delay in the C$_\text{WT}$ strain from this analysis, which cannot be measured directly in experiments.
To this end, we modelled the five strains in our simulations by setting $P_\text{COL}$ and $P_\text{REP}$ to the corresponding values (see Table~S\ref{tab:strains}): $P_\text{COL}=20$ and $P_\text{REP}=0$ for C$_\text{WT}$, $P_\text{COL}=20$ and $P_\text{REP}=55$  for C$_\text{REP1}$, $P_\text{COL}=20$ and $P_\text{REP}=13$  for C$_\text{REP2}$,  $P_\text{COL}=0$ and $P_\text{REP}=55$ for the S$_\text{REP1}$ strain, and  $P_\text{COL}=0$ and $P_\text{REP}=13$ for the S$_\text{REP2}$ strain. For each strain, we simulated \realisations realisations for different ssDNA production rates ($\alpha_D$) to investigate the role of this novel regulatory component. The results of these simulations are depicted in the form of \textit{cea-cel} delay-time histograms in Fig.~S\ref{fig:hist_alphaD}, in which we, for a clear and concise discussion, only depict the results for the C$_\text{REP1}$, S$_\text{REP1}$, and C$_\text{WT}$ strains.

%In order to understand why we find a cea-cel delay in the S$_\text{REP1}$ strain, but not in the C$_\text{REP1}$ strain, we employed the above simulation scheme again to investigate how different rates of ssDNA production affect the delay distribution.
%, and to support our hypothesis that ssDNA crucially affects the delay.  
%Please note that we do not investigate different rates of sRNA production at this point, as the sRNAs are equally present in all strains. 
%Moreover, we did not depict plots for the S$_\text{REP1}$ strain in Fig.~S\ref{fig:hist_alphaD}, as it does not produce and accumulate ssDNA.

As we showed in section~\ref{sec:biosys}, only two factors can in principle be responsible for the different \textit{cea-cel} delays: The total plasmid copy number, and the production of ssDNA.
To separately study the influence of total plasmid copy number in the strains, we first analyse the case of no ssDNA production ($\alpha_{D} = 0$). The delay-time histograms of the three strains for this case are depicted in the first column of Fig.~S\ref{fig:hist_alphaD}.
Comparing the histograms, we find that the total plasmid copy number has a significant effect: The \textit{cea-cel} delay distribution of the C$_\text{REP1}$ strain (75 plasmids in total) has a shorter tail and a more pronounced peak at short lysis times compared to the distribution of the S$_\text{REP1}$ strain (55 plasmids in total), and the C$_\text{WT}$ strain (with only 20 plasmids) shows almost no lysis at all. 
% For the C$_\text{REP2}$ strain (33 plasmids in total), almost two thirds of the cells do not even lyse at all.
The average delay-time of the S$_\text{REP1}$ strain (68 minutes) is in good agreement with experimental values (see Fig.~2E). For the other strains, however, the results do not match: The C$_\text{REP1}$
% and C$_\text{REP2}$
strain has a mean delay time of 24.1%  and 128.5 
minutes, which is significantly larger than the value we find in our experiments. 
Our experiments also find that the wild-type indeed does lyse after SOS responses (see Fig.~S8), while the histogram of C$_\text{WT}$ predicts no lysis. 
% The S$_\text{REP1}$ shows almost no lysis events, which is the case only for about 66\% of the cells in the experiment.
These results for $\alpha_{D} = 0$ show that the plasmid copy number does not suffice to explain the quantitative  and (for the wild-type) qualitative behaviour found in our experiments. However, it already accounts for significant differences in the \textit{cea-cel} delay-time distributions between the strains.

Before we study the additional effects of ssDNA, we discuss the origin of these differences in the \textit{cea-cel} delay-time distributions. To this end, we consider the time evolution of the average levels of free CsrA and long mRNA, which are depicted for $\alpha_{D} = 0$ in the first column of Fig.~S\ref{fig:dyn_alphaD}. 
We find for all three strains that, after the initial equilibration, the average number of CsrA molecules remains at a constant value before an SOS response ($t < 200$ min), which results from the interactions of CsrA with all its binding partners.
Before an SOS signal, the three strains exhibit roughly the same average CsrA level. This changes after the response to an SOS signal ($200$ min $ < t < 500$ min), which reduces the average CsrA level in all three strains:
The C$_\text{REP1}$ strain, containing 75 plasmids, has the lowest CsrA levels, whereas the wild-type strain %, in which only 20 plasmids produce CsrA-sequestering components,
with only 20 plasmids has a significantly higher level.

Comparing the CsrA and long mRNA levels with the corresponding \textit{cea-cel} delay-time distributions in Fig.~S\ref{fig:hist_alphaD} shows that the different \textit{cea-cel} delay time distributions are correlated with the average levels of free CsrA (and free long mRNA):
The lower the average CsrA level during a SOS signal, the shorter the average \textit{cea-cel} delay-time, and the narrower the delay-time distribution.
We can explain this correlation in our mathematical model by the fact that long mRNA production increases in form of stochastic bursts during SOS responses (see section~\ref{sec:mathmod}).
The long mRNAs produced during these bursts must first sequester free CsrA, before their abundance is high enough to produce lysis proteins from it.
For the C$_\text{WT}$ strain, the CsrA level during the SOS response is too high to be sequestered enough by stochastic long mRNA bursts (see Fig.~S\ref{fig:dyn_alphaD} and Fig.~S\ref{fig:hist_alphaD}). For the S$_\text{REP1}$ and the C$_\text{REP1}$ strain, however, CsrA levels reach a sufficiently low average abundance during an SOS response that stochastic bursts of free long mRNA are possible, and eventually lysis protein is produced. The average level of free CsrA therefore determines the probability and hence the timing of lysis. 

In Fig.~S\ref{fig:dyn_alphaD} we can also see that the single trajectories of CsrA abundance differ qualitatively between the three strains: The trajectories of the C$_\text{REP1}$ strain show large and abrupt deviations from the mean value, whereas the abundance of CsrA is closer to the mean in the C$_\text{WT}$ strain. The reason for this difference is the plasmid copy number in each strain: The more plasmids with LexA-regulated promoters, the more CsrA-sequestering elements are produced during an SOS response, and thus the more susceptible the system will be to stochastic bursts in the SOS response, increasing the probability of lysis.
%The higher the (average) CsrA level during the SOS signal, the less likely will bursts in long mRNA occur, and hence the less likely are lysis events.
%as can be seen from a comparison of Fig.~S\ref{fig:dyn_alphaD} with Fig.~S\ref{fig:hist_alphaD}: 
%The lower the average CsrA level during a SOS signal, the shorter the average \textit{cea-cel} delay-time, and the narrower the delay-time distribution.
We already discussed in the previous paragraph that the number of plasmids also strongly affects the average level of free CsrA, as more plasmids cause a larger average number of promoters to be unrepressed. 
This effect is due to the fact that the repressor of the ColicinE2 operon, LexA, stochastically binds to and dissociates from the promoter, and thus triggers long mRNA production for short times even in the absence of an SOS signal. 
The more plasmids present, the larger the number of (transiently) derepressed promoters, and hence the more CsrA-sequestering long mRNA the cell contains. These relations explain the differences in the \textit{cea-cel} delay between the three strains.

%Comparing Fig.~S\ref{fig:dyn_alphaD} with Fig.~S\ref{fig:hist_alphaD} also shows that in the $\alpha_D=0$ case of the C$_\text{WT}$ strain the CsrA level during the SOS signal is too high to be sequestered by stochastic long mRNA bursts.

Finally, in order to characterise the additional effect of ssDNA, we consider the plots with ssDNA production (that is, with $\alpha_D > 0$) in Fig.~S\ref{fig:hist_alphaD}.
As the ssDNA production rate $\alpha_D$ increases from 0, the average delay times in the C$_\text{REP1}$ and C$_\text{WT}$ strain decrease, and also several cells in the C$_\text{WT}$ strain lyse (see the C$_\text{WT}$ histogram for $\alpha_D = 1$ in Fig.~S\ref{fig:hist_alphaD}).
We attribute this to the fact that increasing $\alpha_D$ results in lower average CsrA levels in the C$_\text{REP1}$ and C$_\text{WT}$ strain, see Fig.~S\ref{fig:dyn_alphaD}. Consequently, the average \textit{cea-cel} delay-times in Fig.~S\ref{fig:hist_alphaD} decrease, and lysis of C$_\text{WT}$ bacteria becomes possible.
The S$_\text{REP1}$ strain, which contains no ssDNA-producing ColicinE2 plasmid, but only reporter plasmids, is not affected by the increase of this parameter.
%The plots illustrate the important role of ssDNA for the cea-cel delay time in the C$_\text{REP1}$ and the wild-type strain: In Fig.~S\ref{fig:hist_alphaD}, t
The experimentally observed difference in mean delay times of the C$_\text{REP1}$ and S$_\text{REP1}$ strain occur when the ssDNA production rate reaches $\alpha_{D}=7$ (see Fig.~S\ref{fig:hist_alphaD}). 
At this rate, also the C$_\text{WT}$ shows a broad \textit{cea-cel} delay distribution. For the C$_\text{WT}$ strain, we cannot compare the average \textit{cea-cel} delay-time from our simulations with experimental results, but have to use the lysis time.
% difference of the C$_\text{REP1}$ and C$_\text{WT}$ strains (see above). 
For $\alpha_{D}=7$, this difference is in the same order of magnitude as the experimental results (see Fig.~S8).
If the ssDNA production rate becomes too high,
% (see, the $\alpha_{D}=2.5$ case in Fig.~S\ref{fig:hist_alphaD}), 
large fractions of the cell ensemble lyse even in the absence of an SOS signal, which is not seen for the C$_\text{REP1}$ strain in experiments.

%%Fig.~S\ref{fig:dyn_alphaD} shows that higher ssDNA production values result in a lower CsrA abundance during the time when no SOS signal is present. Once the SOS signal is switched on, it thus takes a shorter time to decrease the CsrA abundance to levels that allow long mRNA production. The natural strain differs from the C$_\text{REP1}$ strain in that it has higher CsrA levels. This might come as a surprise, as the C$_\text{REP1}$ strain's additional 55 reporter plasmids do not produce ssDNA. However, the additional plasmids increase the average number of promoters being unrepressed due to noise. Consequently, more CsrA-sequestering long mRNA is produced in the absence of an SOS signal.

Taken together, these results show that the additional sequestration of CsrA by ssDNA is required for cell lysis in the C$_\text{WT}$ strain, and hence necessary to produce the experimentally observed \textit{cea-cel} delays. Therefore, ssDNA plays a key role in the regulation of ColicinE2 release.
%Low production rates ($\alpha_{D} \approx 0.1 $) give a narrower distribution, which is also shifted towards the beginning of the SOS signal. Further increasing the ssDNA production rate causes most of the realisations to produce free long mRNA very shortly after the beginning of the SOS signal, and some even before. For high production rates, the ssDNA sequesters so many CsrA dimers that free long mRNA appears in all realisations even before any SOS signal \todo{(MO:) not realistic, because CsrA highly abundant?}.
%
%As we already identified this parameter range of the ssDNA production rate to be in the limits of natural systems, we find that our theoretical model confirms the hypothesis that ssDNA is capable of shortening or even removing the delay, as we showed for the C$_\text{REP1}$ strain.

\subsection{sRNA knock-out mutants}
In the main text we also discuss experiments with different sRNA knock-out mutant strains, see Fig. S4.
While the two single knock-out cases (no CsrB or no CsrC) are automatically accounted for by the effective sRNA (see the bullet points on sRNA in section~\ref{sec:mathmod}), the special case of the double sRNA knock-out mutant corresponds to setting $\alpha_{S,0} \equiv \alpha_{S,c} \equiv 0$ in eq.~\eqref{eq:rateS} of our model. This means that no sRNA would be produced, which is the main CsrA-sequestering element in our mathematical model. Therefore, our model predicts a large abundance of free CsrA for the double sRNA knock-out case, and hence a significantly larger \textit{cea-cel} delay. However, we do not see this behaviour in our experiments (see Figs.~S2 and S4), which in contrast show shorter average \textit{cea-cel} delay-times in the double knock-out mutant. These experimental results indicate that, in the absence of the two sRNAs, yet unknown regulatory mechanisms become important. 
In the derivation of our mathematical model (see section~\ref{sec:mathmod}), we eliminated any subordinate targets for CsrA, and focussed on the main CsrA-sequestering elements in \textit{E. coli}, the sRNAs CsrB and CsrC~\cite{Gudapaty2001a}. 
Hence, adjusting our model for the double sRNA knock-out mutant would first require to experimentally investigate the detailed interactions and components of the yet unknown regulatory mechanisms, and then to replace the $S$ component and its interactions correspondingly in the model. 
As the double knock-out mutant is not part of our investigation of the \textit{cea-cel} delay-time in the main text, we do not further extend our model for this very special case. 
In all other strains and mutants discussed in the main text and the Supplementary Information, sRNAs are produced and also are the main CsrA regulator. Therefore, the aforementioned differences between theoretical model and experimental observations that arise with the double sRNA knock-out mutant do not affect any statements we derive for the single knock-out or original strains using our model, eqs.~\eqref{eq:rateM}-\eqref{eq:rateD}.

\newpage 

% \printbibliography

 \newpage
 
 \setcounter{figure}{2} % wird unten nochmal verändert!
 \begin{figure}
 	\centering
 	\caption{Biochemical network involved in the post-transcriptional regulation of ColicinE2. The top part shows the complete network, involving all interactions and components considered in our manuscript. This complex description of the network can be reduced to the set of effective interactions shown the lower panel. The derivation of these effective descriptions is given in section~\ref{sec:mathmod} of the theorz part of the Supplementary Information.
 		%	The left side shows the complete system, as it is described idasasdasdn section~\ref{sec:biosys}, the right side shows the simplified system developed in section~\ref{sec:mathmod}. The corresponding quantities are defined and explained in the respective chapters.
 	}
 	\label{fig:biochemNetwork}
 \end{figure}
 \setcounter{figure}{5} % wurde oben schonmal verändert!
 
 \begin{figure}
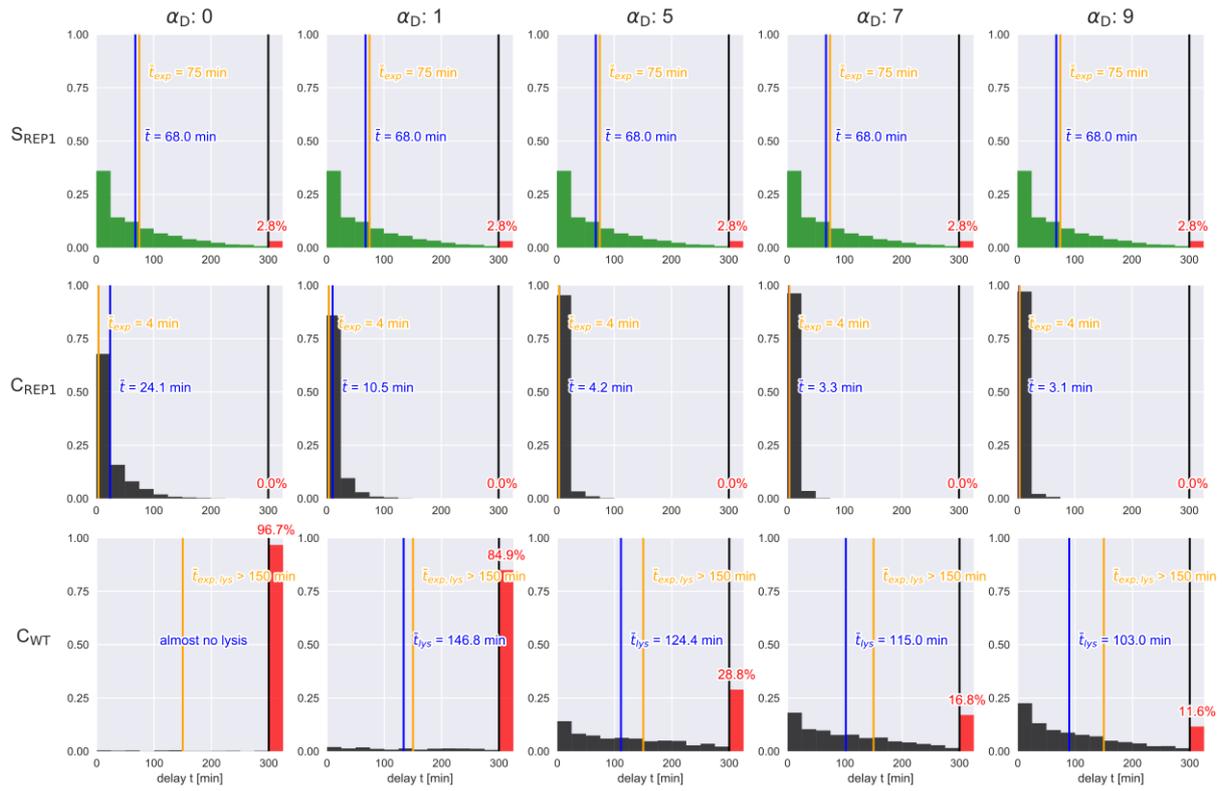

 	\centering
 	\caption{\textit{cea-cel} delay-time distributions and average \textit{cea-cel} delay times for different ssDNA production rates and strains. The S$_\text{REP1}$ strain does not produce ssDNA, and is plotted (in green) only for the purpose of direct comparison with the C$_\text{REP1}$ strain. 
 		If no ssDNA is produced ($\alpha_{D}=0$), we find that the C$_\text{REP1}$ strain shows a broader \textit{cea-cel} delay-time distribution, compared to the cases with ssDNA production. The wild-type strain does not lyse at all in response SOS signal for $\alpha_{D}=0$. If we increase the ssDNA production rate, we find the experimentally observed behaviour that the C$_\text{REP1}$ strain shows no \textit{cea-cel} delays. In the wild type strain, a certain threshold rate of ssDNA production is required to induce a significant level of lysis, emphasising the importance of ssDNA for toxin releases. A comparison of the lysis time ($\bar t_\text{lysis}$) differences between the C$_\text{REP1}$ and C$_\text{WT}$ strains in the model with the lysis time differences in experiments motivates the value of ssDNA production, $\alpha_{D}=1.9$.  Ensemble size: \realisations realisations.}
 	\label{fig:hist_alphaD}	
 \end{figure}
 
 \begin{figure}
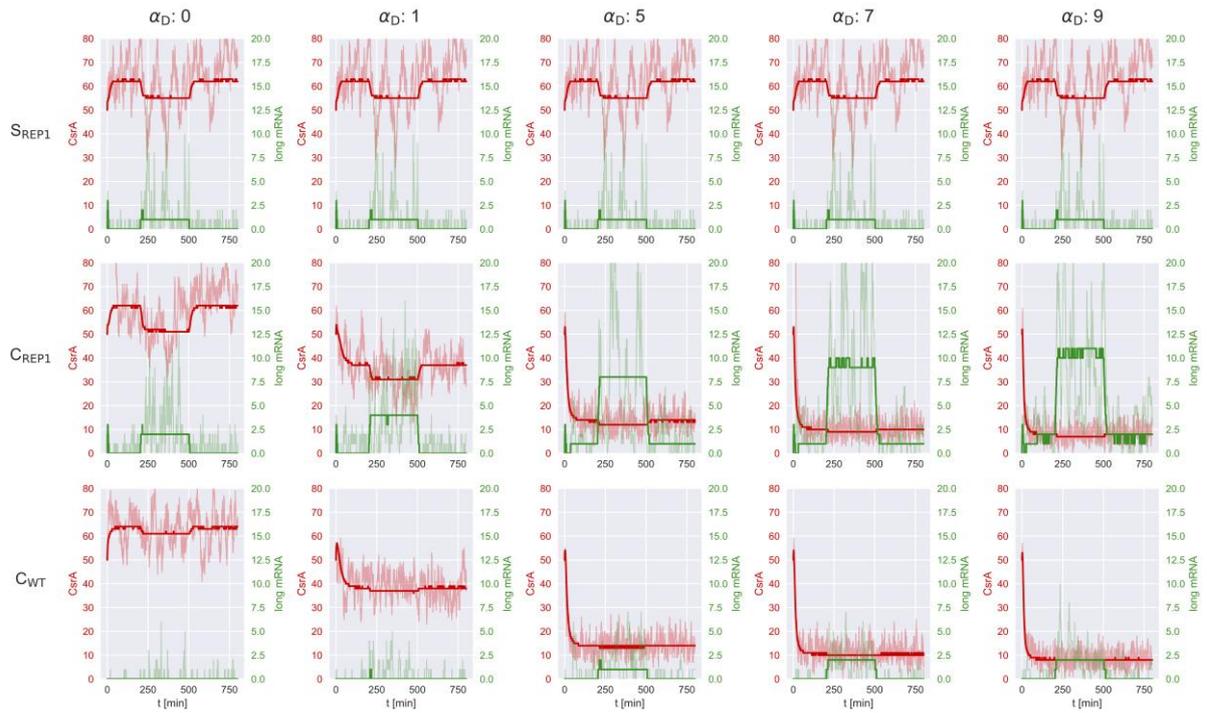

 	\caption{Average of the time evolution of the CsrA and long mRNA abundance for different ssDNA production rates and strains. Between t=200 and t=500, the system is subject to an SOS signal. In all cases, the SOS signal initiates a decrease in CsrA abundance from a previously stable level. This level is determined by the production, binding, and degradation rates of CsrA and its complex partners. As higher levels take longer and are also less likely to decrease to zero, they also directly affect the duration of the average \textit{cea-cel} delay-time. The three plots for $\alpha_D=0$ also show a single trajectory of long mRNA and CsrA in light green and light red, respectively. Ensemble size: \realisations realisations.}
 	\label{fig:dyn_alphaD}	
 \end{figure}


\begin{thebibliography}{10}

\bibitem{Janion2008}
Celina Janion.
\newblock {Inducible SOS Response System of DNA Repair and Mutagenesis in
  Escherichia coli}.
\newblock 4(6):338--344, 2008.

\bibitem{Shimoni2009}
Yishai Shimoni, Shoshy Altuvia, Hanah Margalit, and Ofer Biham.
\newblock {Stochastic Analysis of the SOS Response in Escherichia coli}.
\newblock 4(5), 2009.

\bibitem{Yang2010}
Tsung~Y. Yang, Yun~M. Sung, Guang~Sheng Lei, Tony Romeo, and Kin~F. Chak.
\newblock {Posttranscriptional repression of the cel gene of the ColE7 operon
  by the RNA-binding protein CsrA of Escherichia coli}.
\newblock 38(12):3936--3951, 2010.

\bibitem{Timmermans2010}
Johan Timmermans and Laurence {Van Melderen}.
\newblock {Post-transcriptional global regulation by CsrA in bacteria}.
\newblock 67(17):2897--2908, 2010.

\bibitem{Weilbacher2003}
Thomas Weilbacher, Kazushi Suzuki, Ashok~K. Dubey, Xin Wang, Seshigirao
  Gudapaty, Igor Morozov, Carol~S. Baker, Dimitris Georgellis, Paul Babitzke,
  and Tony Romeo.
\newblock {A novel sRNA component of the carbon storage regulatory system of
  Escherichia coli}.
\newblock 48(3):657--670, 2003.

\bibitem{Mader2015}
Andreas Mader, Benedikt von Bronk, Benedikt Ewald, Sara Kesel, Karin Schnetz,
  Erwin Frey, and Madeleine Opitz.
\newblock {Amount of Colicin Release in Escherichia coli Is Regulated by Lysis
  Gene Expression of the Colicin E2 Operon}.
\newblock 10(3), 2015.

\bibitem{Gudapaty2001a}
Seshagirirao Gudapaty, Kazushi Suzuki, Xin Wang, Tony Romeo, X~I~N Wang, and
  Paul Babitzke.
\newblock {Regulatory Interactions of Csr Components : the RNA Binding Protein
  CsrA Activates csrB Transcription in Escherichia coli Regulatory Interactions
  of Csr Components : the RNA Binding Protein CsrA Activates csrB Transcription
  in Escherichia coli}.
\newblock 2001.

\bibitem{Cascales2007}
Eric Cascales, Susan~K Buchanan, Denis Duché, Colin Kleanthous, Roland
  Lloubès, Kathleen Postle, Margaret Riley, Stephen Slatin, and Danièle
  Cavard.
\newblock {Colicin biology.}
\newblock 71(1):158--229, 2007.

\bibitem{Schwarz2016}
Matthias Lechner, Mathias Schwarz, Madeleine Opitz, and Erwin Frey.
\newblock {Hierarchical Post-transcriptional Regulation of Colicin E2
  Expression in Escherichia coli}.
\newblock 12(12):e1005243, dec 2016.

\bibitem{Ronen2002}
Michal Ronen, Revital Rosenberg, Boris~I Shraiman, and Uri Alon.
\newblock {Assigning numbers to the arrows: parameterizing a gene regulation
  network by using accurate expression kinetics.}
\newblock 99(16):10555--10560, 2002.

\bibitem{Krishna2007}
Sandeep Krishna, Sergei Maslov, and Kim Sneppen.
\newblock {UV-Induced Mutagenesis in Escherichia coli SOS Response: A
  Quantitative Model}.
\newblock 3(3), 2007.

\bibitem{Legewie2008}
Stefan Legewie, Dennis Dienst, Annegret Wilde, Hanspeter Herzel, and Ilka~M
  Axmann.
\newblock {Small RNAs establish delays and temporal thresholds in gene
  expression.}
\newblock 95(7):3232--3238, 2008.

\bibitem{Levine2008}
Erel Levine and Terence Hwa.
\newblock {Small RNAs establish gene expression thresholds}.
\newblock 11(6):574--579, dec 2008.

\bibitem{Edwards2011}
Adrianne~N. Edwards, Laura~M. Patterson-Fortin, Christopher~A. Vakulskas,
  Jeffrey~W. Mercante, Katarzyna Potrykus, Daniel Vinella, Martha~I. Camacho,
  Joshua~A. Fields, Stuart~A. Thompson, Dimitris Georgellis, Michael Cashel,
  Paul Babitzke, and Tony Romeo.
\newblock {Circuitry Linking the Csr and Stringent Response Global Regulatory
  Systems}.
\newblock 80(6):1561--1580, 2011.

\bibitem{Yakhnin2012}
Helen Yakhnin, Alexander~V Yakhnin, Carol~S Baker, Elena Sineva, Igor Berezin,
  Tony Romeo, and Paul Babitzke.
\newblock {Complex regulation of the global regulatory gene csrA: CsrA-
  mediated translational repression, transcription from five promoters by
  E$\sigma$70 and E$\sigma$S , and indirect transcriptional activation by
  CsrA}.
\newblock 81(3):689--704, 2012.

\bibitem{Megerle2008}
Judith~A Megerle, Georg Fritz, Ulrich Gerland, Kirsten Jung, and Joachim~O Ra.
\newblock {Timing and Dynamics of Single Cell Gene Expression in the Arabinose
  Utilization System}.
\newblock 95(August), 2008.

\bibitem{Gillespie1977}
Daniel~T Gillespie.
\newblock {Exact stochastic simulation of coupled chemical reactions}.
\newblock 81(25):2340--2361, 1977.

\end{thebibliography}
\end{document}